\documentclass[twocolumn,prl,aps,showpacs]{revtex4}
\usepackage{graphicx}
\usepackage{dcolumn}
\usepackage{bm}

\begin{document}

\title{\textbf{Direct Measurement of the Spin Hamiltonian and
Observation of \\ Condensation of Magnons in the 2D Frustrated
Quantum Magnet Cs$_2$CuCl$_4$}}
\author{R. Coldea$^{1,2,3}$, D.\ A. Tennant$^{1,3}$, K. Habicht$^4$,
P. Smeibidl$^4$, C. Wolters$^5$ and Z. Tylczynski$^6$}
\affiliation{$^1$Oxford Physics, Clarendon Laboratory, Parks Road,
Oxford OX1 3PU, United Kingdom\\
$^2$Solid State Division, Oak Ridge National Laboratory, Oak
Ridge, Tennessee 37831-6393\\
$^3$ISIS Facility, Rutherford Appleton Laboratory, Chilton, Didcot
OX11 0QX, United Kingdom \\
$^4$Hahn-Meitner Institut, BENSC, D-14109 Berlin, Germany\\
$^5$NHMFL, Florida State University, Tallahassee, Florida 32306\\
$^6$Institute of Physics, Adam Mickiewicz University, Umultowska
85, 61-614 Poznan, Poland}
\date{\today}

\begin{abstract}
We propose a method for measuring spin Hamiltonians and apply it
to the spin-1/2 Heisenberg antiferromagnet Cs$_2$CuCl$_4$, which
shows a 2D fractionalized RVB state at low fields. By applying
strong fields we fully align the spin moment of Cs$_2$CuCl$_4$
transforming it into an effective ferromagnet. In this phase the
excitations are conventional magnons and their dispersion relation
measured using neutron scattering give the exchange couplings
directly, which are found to form an anisotropic triangular
lattice with small Dzyaloshinskii-Moriya terms. Using the field to
control the excitations we observe Bose condensation of magnons
into an ordered ground state.
\end{abstract}

\maketitle



Understanding strongly correlated physics poses formidable
mathematical difficulties and in only a few exceptional cases has
the full many-body quantum problem been solved. Although theory
takes the Hamiltonian ($\mathcal{H}$) as its starting point
linking experimental data to this is often not possible. A method
for measuring $\mathcal{H}$ directly would bridge this gap to
theoretical approaches and in addition reveal the essential
ingredients from which exotic quantum states emerge. Motivated by
this we combine neutron scattering with high magnetic fields and
make just such a determination of $\mathcal{H }$\ taking the
remarkable quantum magnet Cs$_2$CuCl$_4$ as a subject. We base our
approach on overcoming spin couplings using large fields thus
transforming the system into an effective ferromagnet, an easily
solvable state. In addition we explore how the ordered ground
state evolves with lowering field and interpret the results in the
framework of Bose-Einstein condensation (BEC) of magnons.

The insulating magnet Cs$_{2}$CuCl$_{4}$ is an ideal subject for
two reasons: First, its relatively weak ($\sim 4$ K) couplings can
be overcome by current fields (at 8.44~T), and second, it shows
highly unusual strongly correlated properties \cite{Coldea01}.
Among the most fascinating are a low-field dynamics dominated by
2D highly dispersive continua characteristic of fractionalization
of spinwaves into spin-1/2 spinons, exceptionally strong quantum
renormalizations, and an unexplained $T=0$ disordered phase
induced by weak fields along $b$ and $c$. Although Anderson first
proposed a 2D fractionalized state in 1973 (the resonating valence
bond state), the essential conditions for its existance have
remained highly contentious \cite{Anderson73}. In light of this
establishing what the special ingredients are in the Hamiltonian
of Cs$_2$CuCl$_4$ is therefore very important.

The origins of strongly correlated phases lie in the uncertainty
principle. For quantum magnets uncertainty is embedded in the
noncommutation of the spin vector components
${\bm S}=\{S^{x},S^{y},S^{z}\}$ and the ``true'' direction of ${\bm S}$
cannot be known. For spins ${\bm S_{R}}$ on a lattice ${\bm R}$
coupled by the Heisenberg exchange Hamiltonian
\begin{equation}
\mathcal{H}=\frac{1}{2}\sum_{{\bm R},{\bm\delta }}J_{\bm\delta
}{\bm S}_{\bm R}\cdot {\bm S}_{\bm{R}+\bm{\delta}}-g\mu
_{B}BS_{\bm R}^{z} \label{Hamiltonian}
\end{equation}
the energy depends simultaneously on all three
noncommuting components of each ${\bm S_{R}}$ (${\bm\delta }$ is a
vector between sites and the last term an attendant magnetic
field). Quantum uncertainty appears as a \emph{kinetic} term
($S_{\bm R}^{+}S_{\bm R+\bm\delta }^{-}+S_{\bm R}^{-}S_{\bm
R+\bm\delta }^{+}$) in the action on the spins which is most
extreme for spin-1/2 where it flips pairs of spins \textit{e.g.}
$\uparrow \downarrow $ to $\downarrow \uparrow $, and the magnet
fluctuates between many spin configurations. Semiclassically this
kinetic action correlates particle motions (and creation) with
others and can be so strong that new phases emerge
as in Cs$_{2}$CuCl$_{4}$.

When large enough, the field $B$ in (1) prevails over the exchanges and the
unique situation arises where the ground state of $\mathcal{H}$ is known and
the one-particle excited states are exactly solved. The ground
state consists of \emph{all spins up}, which we denote 
$\Psi _{0}=|0\rangle ,\;E_{0}=-Ng\mu _{B}BS+N/2\sum_{\bm\delta
}J_{\bm\delta
}S^{2}$, 
which is indeed that of a ferromagnet. There are $N$ orthonormal
states with a single spin flip $\psi_{\bm R}=S_{{\bm
R}}^{-}/\sqrt{2S}|0\rangle$ corresponding to all sites ${\bm R}$.
When $\mathcal{H}$ acts on $\psi _{\bm R}$, it generates only
other such one-spin-flip states because the total spin
$S_{T}^{z}=\sum_{{\bm R}}S_{{\bm R}}^{z}$ is a
constant-of-the-motion for $\mathcal{H}$. Because the Hamiltonian
is invariant upon translation plane-wave states are diagonal:
$\psi _{\bm k}\equiv N^{-1/2}\sum_{\bm R}\exp(-i{\bm k}\cdot {\bm
R})\psi _{
\bm R}$ 
where $\mathcal{H}\psi _{\bm k}=E({\bm k})\psi _{\bm k}$. The kinetic term
causes hopping of these spin-flips through the lattice and the energy
eigenvalues (for spins $S$=1/2) are
\begin{equation}
E({\bm k})=E_{0}+g\mu _{B}B-J_{\bm 0}+J_{\bm k},\;J_{\bm
k}={\frac{1}{2}} \sum_{\bm\delta }J_{\bm\delta }e^{i {\bm k} \cdot
{\bm\delta }}
\end{equation}
so that the one-spin-flip excitations disperse relative to the
ground state with the relation, $\hbar \omega _{\bm k}=E({\bm
k})-E_{0}=g\mu _{B}B-J_{\bm 0}+J_{\bm k}$, which is a constant
term plus $J_{\bm k}$, \emph{the Fourier transformed exchange
couplings}. These excitations are the familiar quantized harmonic
spin-wave modes, \emph{magnons}, which carry $\Delta S^{z}=-1$ and
have Bose statistics. Since neutrons are spin-1/2 particles they
scatter only by changing the total spin by $\Delta S=0,\pm 1$. For
a system prepared in the fully-aligned state $|0\rangle $ neutrons
can scatter inelastically only by exciting a \emph{single magnon}
through the matrix element $|\langle \psi _{\bm k}|S_{\bm
k}^{-}|0\rangle |^{2}$ where $S_{\bm k}^{-}=N^{-1/2}\sum_{\bm
R}\exp (-i{\bm k}\cdot {\bm R})S_{\bm R}^{-}$. $J_{\bm k}$ and so
$J_{\bm\delta }$ of Eq.\ (\ref{Hamiltonian}) can then be found
from the measured dispersion $\hbar \omega _{\bm k}$.

\begin{figure}[h]
\begin{center}
\includegraphics[width=8cm,bbllx=93,bblly=232,bburx=537,
bbury=653,angle=0,clip=]{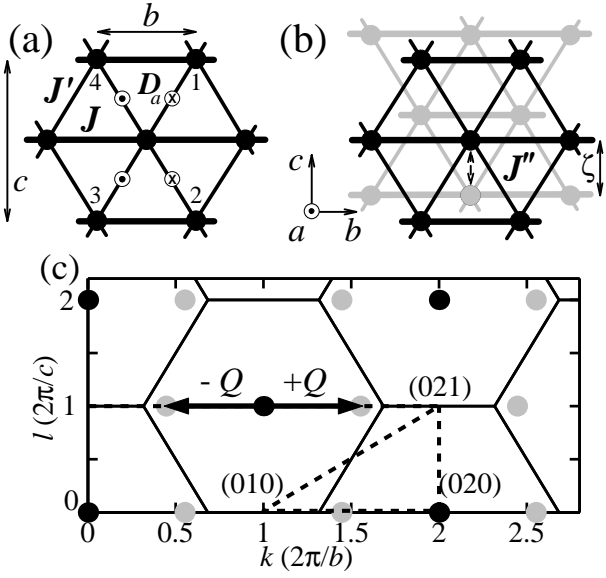} 
\caption{\label{fig1}(a) Magnetic couplings in a 2D triangular
layer in Cs$_{2}$CuCl$_{4}$: strong bonds $J$ (heavy lines
$\parallel b$), smaller frustrating zig-zag bonds $J^{\prime}$
(thin lines). $D_a$ are the Dzyaloshinskii-Moriya (DM) couplings
along the zig-zag bonds; the signs $\otimes,\odot$ refer to
interactions originating at the central spin ${\bm S_{\bm R}}$,
see Eq.\ (\protect\ref{DMhamiltonian}).
(b) Odd (black) and even (grey) layers are stacked successively
along $a$ (inter-layer spacing $a/2$) with an offset $\zeta$=0.34
along $c$. $J^{\prime\prime}$ (dashed arrow) is the
nearest-neighbor inter-layer exchange. (c) ($b^*,c^*$) reciprocal
plane showing the near-hexagonal Brillouin zones (thin lines) of
the triangular lattice in (a).
Black points are zone centers ($\bm \tau$) and grey points at $\bm
\tau \pm \bm Q$ mark dispersion gap minima in the saturated phase
($B>B_C$) and where Bragg peaks appear below $B_C$.}

\end{center}
\end{figure}

The crystal structure of Cs$_{2}$CuCl$_{4}$ is orthorhombic
(\emph{Pnma}) with lattice parameters at 0.3 K of $a$=9.65 \AA ,
$b$=7.48 \AA , and $c$=12.35 \AA . The magnetic $S=1/2$ Cu$^{2+}$
ions are situated within distorted CuCl$_4^{2-}$ tetrahedra.
Layers ($bc$ plane) of these tetrahedra are separated by Cs$^{+}$
ions and are stacked with an offset giving the structure
illustrated in Fig.\ \ref{fig1}(a). The strongly correlated
physics derives from the ``isosceles'' triangular lattice
arrangement of spins in the layers with antiferromagnetic exchange
paths $J\equiv J_{[0,1,0]}$ and $J^{\prime }\equiv J_{[0,1/2,\pm
1/2]}$. The triangular geometry allows a large configuration space
for fluctuations and is presumably crucial to fractionalization.

To measure $\hbar \omega _{\bm k}$ the V2 cold-neutron triple-axis
spectrometer at the BER-II reactor at HMI in Berlin was used. A
large (3.6 g) high-quality single crystal of Cs$_{2}$CuCl$_{4}$
was mounted in the (0,$k$,$l$) scattering plane on a dilution
refrigerator insert with base temperature of 50 mK. The VM-1
cryomagnet provided fields up to 14.5 T along $a$. The
spectrometer was configured with a vertically-focused
monochromator (PG002) and a horizontally focused PG002 analyzer to
select scattered neutrons with fixed $k_{f}=1.2$ or 1.35
{\AA}$^{-1}$.

\begin{figure}[h]
\begin{center}
\includegraphics[width=8cm,bbllx=18,bblly=70,bburx=559,
bbury=798,angle=0,clip=]{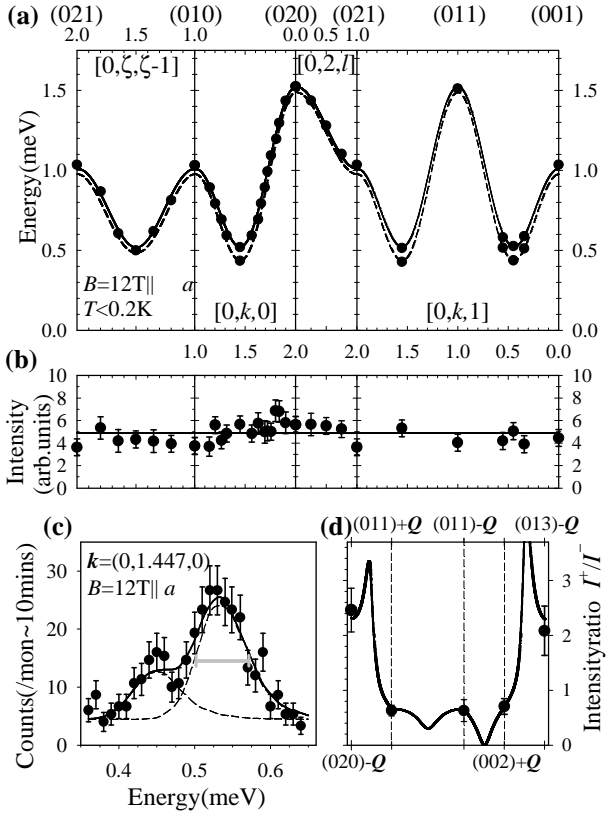} 
\caption{\label{fig2} (a) Magnetic dispersion relations in the
saturated phase ($B$=12 T$\parallel$$a$, $T<$0.2 K) along symmetry
directions in the 2D plane (heavy dashed lines in Fig.\
\protect\ref{fig1}(c)). Solid and dashed lines are fits to Eq.\
(\protect\ref{dispersions}) with parameters in Table I. (b)
Observed integrated inelastic intensity compared with predictions
for the fully-polarized eigenstate (solid line). (c) Excitations
lineshape observed along a constant-wavevector scan at the minimum
gap $\bm{k}$=(0,1.447,0). Solid line is a fit to Eqs.\
(\protect\ref{dispersions}-\protect\ref{intensities}) convolved
with the instrumental resolution (horizontal grey bar indicates
the full-width-half-maximum of the energy resolution).(d) Relative
intensity of the two magnon modes compared with Eq.\
(\ref{intensities}) (solid line).}
\end{center}
\end{figure}

A magnetic field of 12 T, much larger than the saturation field (8.44 T),
was used to open a significant energy gap of 0.435(8) meV to the first
excited states. Temperatures below 200 mK ensured that the
thermally introduced population of spin flips was less than 1 per $10^{11}$
spins. No magnetostructural distortions were observed and the origin of
superexchange in high-energy electronic bonds means that the coupling
constants are unperturbed by the field. Only one magnon scattering events
were observed and their energy and wavevector dependence mapped
out. Fig.\ \ref{fig2}(c) shows a typical scan. Two resolution limited peaks
are seen separated by a small energy of 0.084(2) meV; this
splitting is due to an additional anisotropy as explained below.

The measured one-magnon dispersion relations are graphed in Fig.\
\ref{fig2}(a). The considerable dispersion along both [0$k$0] and
[00$l$] in the $bc$ plane indicates strong 2D character. The
overall dispersion follows $\hbar \omega _{\bm k}$ with $J_{\bm
k}=J\cos (2\pi k)+2J^{\prime }\cos (\pi k)\cos (\pi l)$ where $
J=0.374(5)$ meV and $J^{\prime }=0.128(5)$ meV, the couplings in
Fig.\ \ref {fig1}(a) (${\bm k}=(h,k,l)$ is expressed in units of
($2\pi /a,2\pi /b,2\pi /c)$). The small splitting into two magnon
branches is characterized well by modified dispersions $\hbar
\omega _{\bm k}^{\pm }=g\mu _{B}B-J_{\bm0}+J_{\bm k}\pm D_{\bm
k},D_{\bm k}=2D_{a}\sin (\pi k)\cos (\pi l)$ with $D_{a}=0.020(2)$
meV. This is surprising because whereas $J_{\bm k}$ is a sum of
cosine terms, $D_{\bm k}$ is \emph{sinusoidal}. The physical
meaning of this is that a left moving magnon (of a certain type)
has different energy from a right moving one $\hbar \omega_k^{\pm
}\neq \hbar \omega _{-k}^{\pm }$; the two magnon branches actually
cross over at $k=0$, $\hbar \omega _{k}^{\pm}=\hbar
\omega_{-k}^{\mp}$, and such a situation can come about only if an
exchange with a sense of direction is present.

Dzyaloshinskii and Moriya (DM) \cite{Moriya60} proposed just such an
exchange interaction many years ago. They showed that spin-orbit couplings
in the superexchange can generate a coupling of the form ${{\bm D}_{ij}}%
\cdot ({\bm S}_{i}\times {\bm S}_{j})$. Their interaction is of higher-order
and therefore much weaker than Heisenberg exchange and can occur only when
the superexchange pathways do not have centers of inversion which is indeed
the case in Cs$_{2}$CuCl$_{4}$.

In the ordered structure (below $T_{N}$=620~mK and $B=0$) the
moments lie almost within the $bc$ plane which would happen if the
most important ${\bm D}_{ij}$ vector in Cs$_{2}$CuCl$_{4}$ was
directed along the $a$-axis, and the observed wavevector
dependence, $\sin (\pi k)\cos (\pi l)$, of $D_{\bm k}$ suggests
that the important DM interaction is along the same zig-zag bonds
as $J^{\prime }$ in the 2D planes. Considering these bonds only,
and making the approximation that ${\bm D}^{\pm }=(\pm
D_{a},0,0)\equiv {\bm D} _{a}^{\pm }$ we obtain using symmetry
\begin{equation}
\mathcal{H}_{DM}^{\pm }\!\!=\!\frac{1}{2}\!\sum_{\bm R}{\bm
D}_{a}^{\pm }\cdot {\bm S_{\bm R}}\times \lbrack -{\bm
S_{\bm{R}+\bm{\delta}_{1}}}-
{\bm{S}_{\bm{R}+\bm{\delta}_{2}}}+
{\bm{S}_{\bm{R}+\bm{\delta}_{3}}}+{\bm{S}_{\bm{R}+\bm{\delta}_{4}}}]
\label{DMhamiltonian}
\end{equation}
where the labels $\bm\delta _{1-4}$ refer to Fig.\ \ref{fig1}(a)
and the $ \pm $ has been introduced because there are two distinct
layers shown in Fig.\ \ref{fig1}(b) which are inverted versions of
each other with DM vectors pointing in opposite directions. Like
the Heisenberg coupling this DM interaction also conserves
$S_{T}^{z}$ and plane-wave solutions remain diagonal;
$\mathcal{H}_{DM}^{\pm }\psi _{\bm k}=\pm D_{\bm k}\psi _{\bm k}$
where $D_{\bm k}=2D_{a}\sin (\pi k)\cos (\pi l)$ as observed. The
DM interaction then explains the observed sinusoidal components of
$\hbar \omega _{\bm k}^{\pm }$ and the fact that there are two
modes - one for each type of layer.

\begin{table}[tbp]
\caption{Hamiltonian parameters ($B>B_C$)(see text) versus the quantum
renormalized ($B=0$) parameters from \cite{Coldea01}.}
\begin{ruledtabular}
\begin{tabular}{cccc}
Parameter&$B>B_C$&$B=0$&Renormalization\\
\hline
$J$ (meV) & 0.374(5) & 0.62(1) & 1.65(5)\\
$J'$ (meV) & 0.128(5) & 0.117(9) & 0.91(9)\\
$J''$ (meV) & 0.017(2) & - & -\\
$D_a$ (meV) & 0.020(2) & - & -\\
$\epsilon$ (rlu) & 0.053(1) & 0.030(2) & 0.56(2)\\
\end{tabular}
\end{ruledtabular}
\end{table}

The fact that Cs$_2$CuCl$_4$ orders three dimensionally means that there
\emph{must} be an interaction $J^{\prime\prime}_{\bm k}$ between layers. We
introduce operators $a_{\bm k}^\dagger $ and $b_{\bm k}^\dagger$ that create
the two types of magnons on the different layers. The full Hamiltonian with
DM and interlayer couplings is:
\[
\mathcal{H}\!=\!\!\sum_{\bm k} \! \left[\!
\begin{array}{cc}
a_{\bm{k}}^{\dag} & b_{\bm{k}}^{\dag}
\end{array}
\! \right] \! \!\left[
\begin{array}{cc}
{\hbar\Omega_{\bm{k}}-J^{\prime\prime}_{\bm{0}}}+D_{\bm{k}} &
J^{\prime\prime}_{-\bm{k}} \\
J^{\prime\prime}_{\bm{k}} &
{\hbar\Omega_{\bm{k}}-J^{\prime\prime}_{\bm{0}}}-D_{\bm{k}}
\end{array}
\! \right] \! \! \left[ \!
\begin{array}{c}
a_{\bm{k}} \\
b_{\bm{k}}%
\end{array}
\! \right] \nonumber
\]
where $\hbar \Omega_{\bm{k}}=g\mu_BB-J_{\bm 0}+J_{\bm k}$ is the magnon
dispersion for $D_{\bm k} = J^{\prime\prime}_{\bm k} = 0$. Diagonalizing
this Hamiltonian gives the new dispersion relations
\begin{equation}
\hbar\omega_{\bm{k}}^{\pm}=\hbar\Omega_{\bm{k}}-J^{\prime\prime}_{\bm{0}} \pm%
\sqrt{D_{\bm{k}}^2+|J^{\prime\prime}_{\bm{k}}|^2},  \label{dispersions}
\end{equation}
and for the case of interlayer nearest neighbor coupling [see Fig.\ \ref%
{fig1}(b)] ($J^{\prime\prime}_{\bm{k}}$=$J^{\prime\prime}\cos (\pi h)e^{- i
2\pi l \zeta}$) the relative intensity of the two modes is
\begin{equation}
\frac{I^{+}_{\bm{k}}}{I^{-}_{\bm{k}}}=\frac{1+\cos(2\pi l
\zeta)\sin(2\theta_{\bm{k}})} {1-\cos(2\pi l \zeta)\sin(2\theta_{\bm{k}})},
\label{intensities}
\end{equation}
with $\theta_{\bm{k}}=\tan^{-1}[J^{\prime\prime}\cos (\pi
h)/(\sqrt{D_{\bm{k}}^2+|J^{\prime\prime}_{\bm{k}}|^2}+D_{\bm{k}})]$
and where the total inelastic intensity $I^+_{\bm{k}} +
I^-_{\bm{k}}$ is independent of wavevector. Here $\zeta$=0.34 is
the relative offset along $c$ between adjacent layers. Fitting the
above model (Eqs.(\ref{dispersions}) and (\ref{intensities})) to
the data yields the excellent fits shown in Figs.\ \ref
{fig2}(a)-(d) with the fitted parameters listed in the first
column of Table I and $g_a$=2.19(1). The total inelastic intensity
shown in Fig.\ \ref{fig2}(b) is nearly independent of ${\bm k}$ as
predicted. The relative intensity of the two modes (where they
could be resolved) is shown in Fig.\ \ref{fig2}(d). We conclude
that all other couplings in Cs$_2$CuCl$_4$ are much smaller.
Dipolar energies and g-tensor anisotropies are small and neglected
here.

\begin{figure}[h]
\begin{center}
\includegraphics[width=8cm,bbllx=70,bblly=104,bburx=510,
bbury=770,angle=0,clip=]{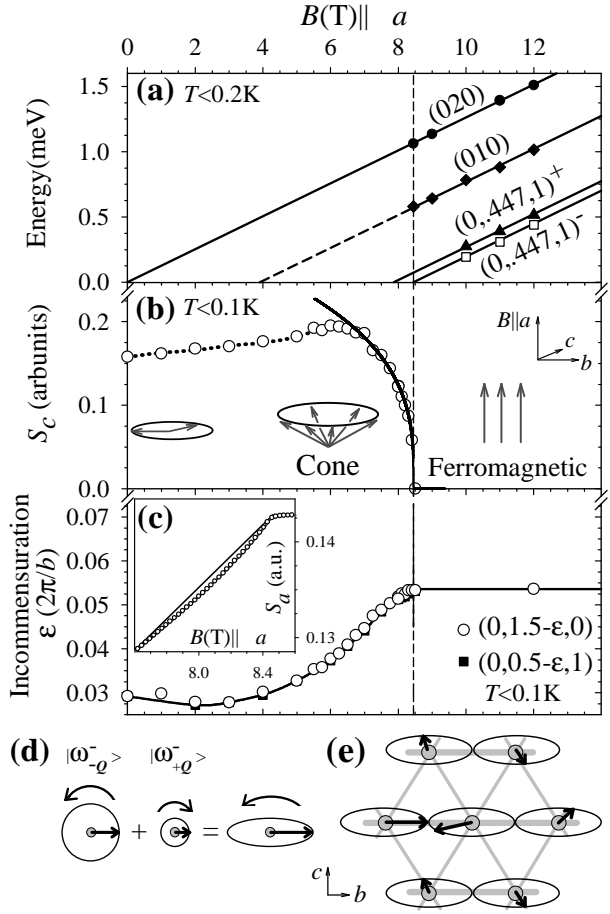} 
\caption{\label{fig3}(a) Magnon energies vs. field in the
saturated phase. Solid lines are fits to a linear behavior as
expected for $\Delta S^z=-1$ eigenstates with $g$-factor
$g_a$=2.18(1). (0,0.447,1)$^{\pm}$ label the two magnon modes
resolved at the minimum gap in scans such as in Fig.\
\protect\ref{fig2}(c). (b) Amplitude of perpendicular ordered
moment $S_c$ in the cone phase vs. field. Solid line is a
power-law fit. (c) Incommensuration ($\epsilon$=$Q$-0.5) vs. field
(solid line is guide to the eye). Inset: magnetization vs. field
\protect\cite{longpaper} ($T$=30 mK) compared with a linear
behavior (solid line). (d) Superposition of contrarotating magnons
$\omega_{\pm \bm Q}^{-}$ \protect\cite{wavefunctions} of different
amplitudes (large and small circle) gives the elliptical order in
$bc$ plane shown schematically for odd layers in (e) (arrows are
ordered spins). Even layers have opposite sense of rotation.}
\end{center}
\end{figure}

Upon decreasing field the magnon energies reduce by the additive
Zeeman term $g\mu _{B}B$ [see Fig.\ \ref{fig3}(a)]. At the
critical field $B_{C}$=8.44(1) T the gap closes at the dispersion
minima $\bm{\tau}\pm \bm{Q}$, $\bm{Q}$=(0.5+$\epsilon
$)$\bm{b}^{\ast }$, $\epsilon $=0.053(1). At those wavevectors
Bragg peaks appear below $B_C$ indicating transverse
(off-diagonal) long-range order. This order is an example of BEC
in a dilute gas of magnons induced by changing the ``chemical
potential'' $|B_C-B|$ \cite{Matsubara56}. The measured spin order
forms an elliptical cone around the field direction $\langle
\bm{S}_{\bm R}\rangle \!=\!\pm \hat{\bm b}S_{b}\cos \bm Q\!\cdot
\!\bm R+\hat{\bm c}S_{c}\sin \bm Q\!\cdot \!\bm R+\hat{\bm
a}S_{a}$
(odd/even $\pm $ layers contrarotate) where $S_{b}>S_{c}$ as
illustrated in Fig.\ \ref{fig3}(e). In fact \emph{this order
corresponds exactly to the simultaneous condensation of
contrarotating magnons} $\omega _{-\bm Q}^{-}$ \emph{and} $\omega
_{+\bm Q}^{-}$ [see Fig.\ \ref{fig3}(d)] \emph{with gap closure
at} $B_{C}$; a mean-field calculation
\cite{Nikuni95,wavefunctions} of this state gives an elliptical
cone with asymmetry $(\cos \theta _{\bm Q}+\sin \theta _{\bm
Q})/(\cos \theta _{\bm Q}-\sin \theta _{\bm Q})$=1.52(6) in
agreement with the observed ratio $S_{b}/S_{c}$=1.55(10) just
below $B_{C}$. The asymmetry is a combined effect of interlayer
coupling $J^{\prime \prime }$ and alternation of $\bm D^{\pm }$
between layers and rapidly decreases as the field is lowered due
to increased inter-particle interactions and fluctuations,
$S_{b}/S_{c}$=1.1(1) below 7 T.

The effect of fluctuations and interactions on the order as field
decreases is quantified in Fig.\ \ref{fig3}(b-c): Fig.\
\ref{fig3}(b) shows the off-diagonal order parameter $S_{c}$.
Close to $B_{C}$ it is described by a power law (solid line)
$S_{c}$$\sim
$$|B_{C}-B|^{\beta }$ with $\beta$=0.33(3), significantly below
the value $\beta$=0.5 expected for mean-field (3D) BEC
\cite{Matsubara56}. The magnetization $S_{a}$ obtained from
susceptibility measurements \cite{longpaper} is plotted in the
inset of Fig.\ \ref{fig3}(c). It shows that the boson density is
not linear versus $|B_{C}-B|$ but rather shows a deviation that
may be logarithmic \cite{Sachdev94}; and finally the wavevector of
the condensate $Q=0.5+\epsilon(B)$ is plotted in Fig.\
\ref{fig3}(c). $Q$ varies strongly with field indicating that
magnon-magnon interactions are important even at low density and
renormalize the condensate wavevector. The above features deviate
significantly from mean-field (3D) behavior \cite{Nikuni00} and
could be associated with the 2D nature of the magnons. In two
dimensions interactions can qualitatively change the scaling
behavior such as by introducing non-linear, log corrections to the
magnetization curve \cite{Sachdev94}.

In summary, we have determined the Hamiltonian of the quasi-2D
quantum magnet Cs$_2$CuCl$_4$ using a new experimental method and
show that it is a 2D anisotropic triangular system. We also
measured transverse (off-diagonal) order with field below
saturation, an example of Bose-Einstein condensation of magnons.
Our methods are general and could be used to reveal exchanges and
quantum renormalizations for systems as diverse as random magnets,
quantum antiferromagnets and spin glasses.

We would like to thank P. Vorderwisch for technical support and
R.A. Cowley, A.M. Tsvelik and F.H.L. Essler for stimulating
discussions. ORNL is managed for the US DOE by UT-Battelle, LLC,
under contract DE-AC05-00OR22725. Financial support was provided
by the EU through the Human Potential Programme under IHP-ARI
contract HPRI-CT-1999-00020.

\begin{thebibliography}{99}
\bibitem{wavefunctions} The two magnon wavefunctions are $|\omega_{-\bm
Q}^{-}\rangle$=$-\cos\theta_{\bm Q}\psi^{odd}_{-\bm Q}$
+$\sin\theta_{\bm Q}\psi_{-\bm Q}^{even}$ and $|\omega_{+\bm
Q}^{-}\rangle$=$-\sin\theta_{\bm Q}\psi_{\bm
Q}^{odd}$+$\cos\theta_{\bm Q}\psi_{\bm Q}^{even}$, where
$\psi_{\pm \bm Q}^{odd/even}$ are plane-wave superpositions of
single spin flip states localized on the odd/even layers.
\end{thebibliography}
\end{document}